\documentclass[sigconf]{acmart}

\usepackage[linesnumbered,ruled,vlined]{algorithm2e}
\usepackage{algorithmic}
\usepackage{footmisc}
    
\AtBeginDocument{%
  \providecommand\BibTeX{{%
    \normalfont B\kern-0.5em{\scshape i\kern-0.25em b}\kern-0.8em\TeX}}}

\setcopyright{acmcopyright}
\copyrightyear{2023}
\acmYear{2023}
\acmDOI{XXXXXXX.XXXXXXX}

\acmPrice{15.00}
\acmISBN{978-1-4503-XXXX-X/18/06}

\begin{document}

\title{UI Semantic Group Detection: Grouping UI Elements with Similar Semantics in Mobile Graphical User Interface}
\author{Shuhong Xiao}
\affiliation{%
  \institution{Zhejiang University}
  \city{Hangzhou}
  \country{China}
  \postcode{310027}
}

\author{Yunnong Chen}
\affiliation{%
  \institution{Zhejiang University}
  \city{Hangzhou}
  \country{China}
  \postcode{310027}
}

\author{Yaxuan Song}
\affiliation{%
  \institution{Zhejiang University}
  \city{Hangzhou}
  \country{China}
  \postcode{310027}
}

\author{Liuqing Chen}
\authornote{Corresponding author: chenlq@zju.edu.cn}
\affiliation{%
  \institution{Zhejiang University}
  \city{Hangzhou}
  \country{China}
  \postcode{310027}
}
\affiliation{%
  \institution{Alibaba-Zhejiang University Joint Research Institute of Frontier Technologies}
  \city{Hangzhou}
  \country{China}
  \postcode{310027}
}

\author{Lingyun Sun}
\affiliation{%
  \institution{Zhejiang University}
  \city{Hangzhou}
  \country{China}
  \postcode{310027}
}
\affiliation{%
  \institution{Alibaba-Zhejiang University Joint Research Institute of Frontier Technologies}
  \city{Hangzhou}
  \country{China}
  \postcode{310027}
}

\author{Yankun Zhen}
\affiliation{%
  \institution{Alibaba Group}
  \city{Hangzhou}
  \country{China}
  \postcode{311121}
}

\author{Yanfang Chang}
\affiliation{%
  \institution{Alibaba Group}
  \city{Hangzhou}
  \country{China}
  \postcode{311121}
}
\begin{abstract}
Texts, widgets, and images on a UI page do not work separately. Instead, they are partitioned into groups to achieve certain interaction functions or visual information. Existing studies on UI elements grouping mainly focus on a specific single UI-related software engineering task, and their groups vary in appearance and function. In this case, we propose our semantic component groups that pack adjacent text and non-text elements with similar semantics. In contrast to those task-oriented grouping methods, our semantic component group can be adopted for multiple UI-related software tasks, such as retrieving UI perceptual groups, improving code structure for automatic UI-to-code generation, and generating accessibility data for screen readers. To recognize semantic component groups on a UI page, we propose a robust, deep learning-based vision detector, UISCGD, which extends the SOTA deformable-DETR by incorporating UI element color representation and a learned prior on group distribution. The model is trained on our UI screenshots dataset of 1988 mobile GUIs from more than 200 apps in both iOS and Android platforms. The evaluation shows that our UISCGD achieves 6.1\% better than the best baseline algorithm and 5.4 \% better than deformable-DETR in which it is based. 
\end{abstract}






\maketitle

\section{Introduction}
\label{sec:introduction}

In the realm of cognitive psychology, humans often tend to organize and group the information they encounter, enabling them to understand and process it more effectively \cite{eysenck2023fundamentals}. This phenomenon is especially pronounced in the design, production, and use of Graphic User Interfaces (GUIs).
In the design process, designers utilize software tools like Figma \cite{figma} to create UI prototypes. The grouping of UI elements are fashioned in the view hierarchy by wrapping basic visual elements within additional containers, serving to manage the broader layout and design style. When it comes to implementation, the grouping of UI elements is handled through HTML tags such as ``div'', which not only enhances code readability and maintainability \cite{xiao2022ui} but also bolsters page loading speed and performance during interactions \cite{chen2023egfe,29}. From a user's standpoint, the grouping of UI elements is accomplished through human visual perception, allowing users to form a holistic understanding of the UI layout and facilitating easier navigation.

The process of grouping UI elements is a crucial component of UI visual intelligence \cite{perceptual_group} and is widely discussed in academic circles pertaining to UI-related software engineering tasks including UI testing, automation and interaction which are the three topics of high interests. For UI testing, elements grouping strengthens the ability of pixel-based non-intrusive approach \cite{RoScript}. By grouping related UI elements into larger components, we can support not only widget-level tests \cite{icon_annotation,27} but examine the interaction among multiple elements \cite{element_interaction}, thereby making automated testing more effective. Automatic GUI production, including prototype design and code generation, also profits from UI elements grouping. Existing methods for design search either perform queries of the entire GUI appearance \cite{UI_design_search,rico} or widgets with regular color and size \cite{28}, while element grouping can fill the gap of components (or module) level searching. When it comes to code generation,  the addition of an element grouping stage \cite{UILM} offers prior structure knowledge, which helps to generate less redundant code, especially for those approaches that only utilize pixel information \cite{sketch2code,mock-up_generation}. Moreover, UI elements grouping improves the interactive experience of the software. For example, screen readers \cite{screen_reader,Screen_Recognition} are designed to help visually impaired users access software by reading out the content based on pre-defined accessibility metadata \cite{accessibility}. GUI elements grouping facilitates the generation of accessibility metadata in higher-order units \cite{perceptual_group} (e.g., components) instead of just widgets.

\begin{figure*}[tbh]
    \centering
    \includegraphics[width=0.95\linewidth]{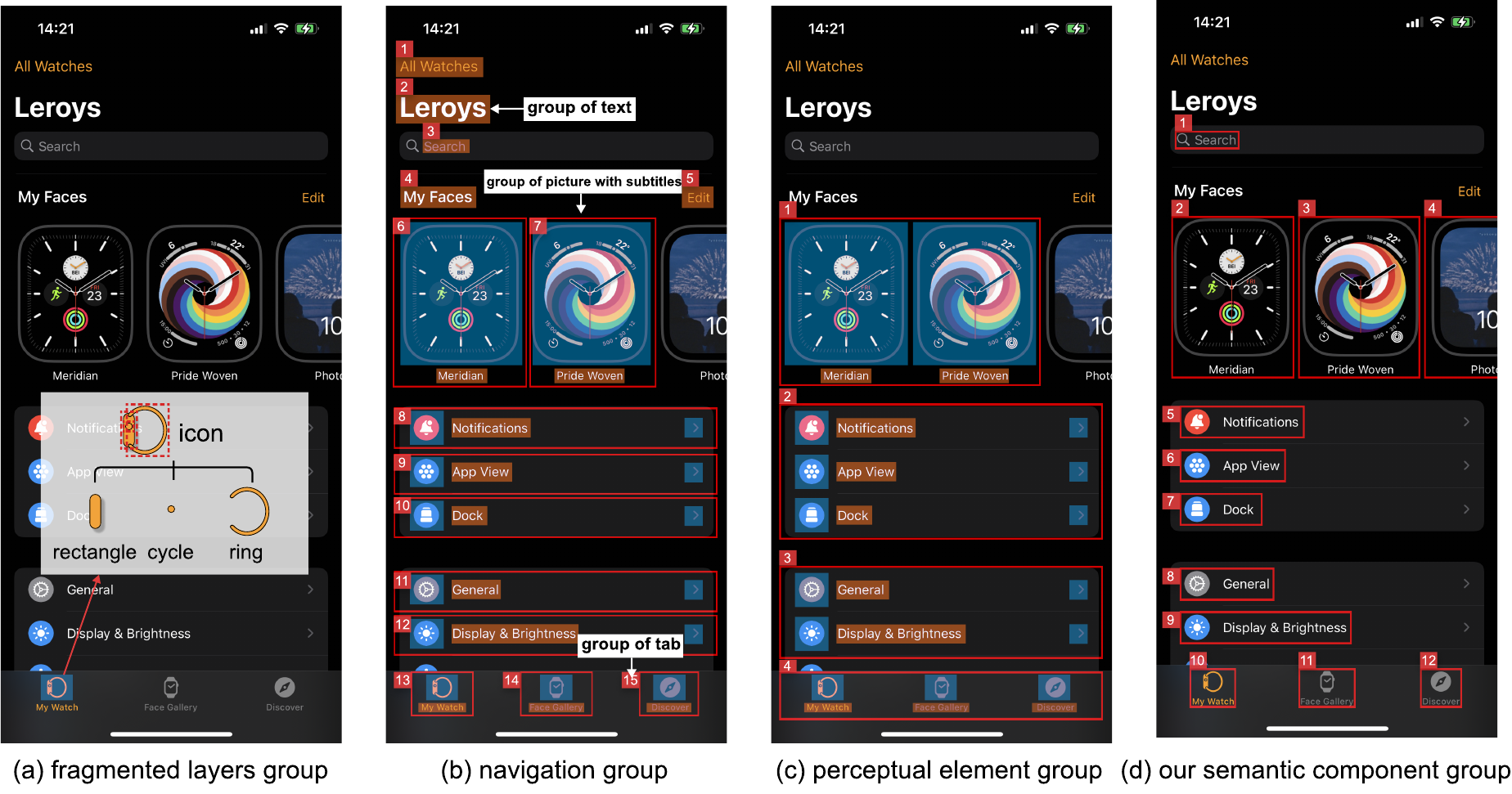}
    \caption{Examples of UI elements groups: (a) fragmented UI layers group; (b) navigation groups for screen reader accessibility; (c) psychologically-inspired perceptual groups; (d) our semantic component groups. Groups are labeled by red bounding boxes.}
    \label{fig1}
\end{figure*}

Despite the extensive application and discussion surrounding the grouping of UI elements, the most pronounced distinction lies in the granularity of these UI groups. The term granularity here refers to the scale of the groups, which can range from broader, perceptually consistent groups to finer, more detailed components. This variance in granularity serves different software UI tasks and there is often a lack of general applicability among them. In Fig. \ref{fig1}, we present three prevalent types of UI groups: fragmented element group, navigation group, and perceptual group, arranged from the smallest to the largest in terms of granularity. The fragmented element group \cite{UILM,chen2023egfe} occurs on the process of automatic front-end code generation from design prototypes. Each fragmented element group encompasses several basic vector shapes that together form a fundamental element of the GUI visual effect. For instance, the ``watch” icon we masked in Fig. \ref{fig1}(a) is actually composed of three basic vector shapes(a circle, an arc, and a rectangle) in the design prototype. With the recognition and organization of such elements through fragmented groups, the automated code generation process can accurately produce the description for visual effect like the ``watch” icon, rather than for each vector shape element. This in turn enhances the quality of the generated code. The fragmented layers group represents the smallest granularity of grouping so far, with its objects so minute that they can be challenging to discern from a UI visual effect. Another significant study, Screen Recognition \cite{Screen_Recognition} forms a navigation group, as shown in Fig. \ref{fig1}(b). This group is utilized to generate the missing accessibility data for UI element objects, specifically aimed at enhancing user experience for visually impaired users. Based on the detection of text and image elements, the groups are formed by pre-defined distance rules, in which case we could see groups in component-level granularity (the group of tab contains icon and text) and element-level granularity (the group of text only). The psychologically-inspired perceptual group by Xie et al. \cite{perceptual_group}, as shown in Fig. \ref{fig1}(c), discusses a section-level group (such as menus, multi-tabs, and cards). Each perceptual group encompasses several UI components with similar structure and function. By segmenting the UI interface into sections, perceptual grouping effectively captures the holistic style of a design element. This comprehensive representation significantly enhances the effectiveness of design search tasks \cite{rico,28}.

In the review of past works on UI element grouping, most efforts \cite{perceptual_group,UILM,Screen_Recognition} have been centered around the detection of single UI objects, such as text, buttons, icons, and check boxes. However, as each UI-related downstream software task relies on a unique form of grouping (Fig. \ref{fig1}), the transition from individual element detection results to final group formations is typically addressed separately. Most approaches rely on rule-based methods at this stage, failing to fully leverage the advantages of deep learning algorithms. On a positive note, the focus on the detection of individual UI object has resulted in a wealth of available datasets that can be directly utilized. However, the dependence on non-intelligent grouping processes has introduced constraints on the effectiveness of these methods. For instance, Chen et al. \cite{chen2023egfe} highlighted that relying exclusively on the identification of individual elements for grouping falls short of achieving the best performance, underscoring the need for more sophisticated methods. To surmount this limitation, we shifted our focus away from the detection of single UI objects and instead commenced with detection at some element groups. Compared to the process from individual elements to groups, transitioning from smaller granularity groups to larger granularity groups is more straightforward. This ease is attributed to the nested nature of UI structure, where larger groups are typically composed of multiple smaller ones that share similar shapes and are arranged according to specific alignment rules. Based on this foundation, we target the most basic UI groups that incorporate both text and image elements. As shown in Fig. \ref{fig1}(d), this is considered a component-level group. Due to the contained text and image elements often being semantically identical or complementary, we term these as UI semantic component groups. Such forms of grouping are very common in UI design. Structurally, these groups are usually defined within the frontend code by a "div" container that sets their boundaries. As illustrated in Fig. \ref{fig2}, leveraging the scalability of iterative grouping, we are able to extend it to a variety of downstream UI tasks, achieving superior performance. Similar to the achievements of \cite{UILM} at the element level with fragmented element groups, our semantic component groups offer robust structural guidance at the component level for UI-to-code generation tasks. Furthermore, our approach inherits the concept of perceptual grouping, utilizing grouping to decipher UI structure. We expose a smaller partition with a singular component on the UI page, which can be further amalgamated into perceptual groups. Unlike the navigation group, our groups harness the corresponding text to facilitate image interpretation, thereby streamlining the acquisition of accessibility data.
\begin{figure*}[tbh]
    \centering
    \includegraphics[width=0.7\linewidth]{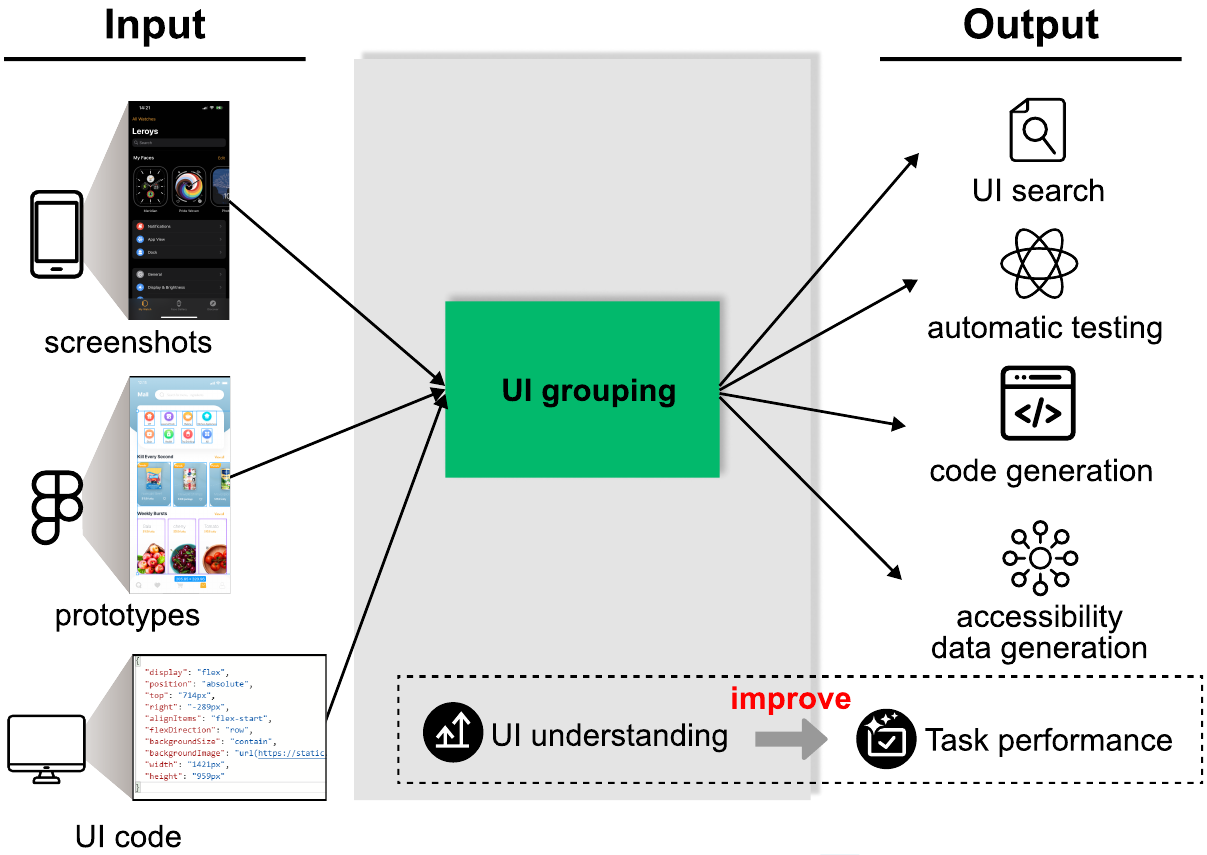}
    \caption{UI grouping for software tasks}
    \label{fig1.2}
\end{figure*}

To obtain semantic component groups, we propose our UI Semantic Component Group Detector (UISCGD) that presents grouping as a one-stage task based on a data-driven approach. We take the enhanced version of Deformable DETR \cite{19} as our baseline detector. It includes features such as muti-scale representation, regional proposal, and an attention mechanism, which show good performance in generic object detection \cite{fatser_rcnn,center_net,small_object_detection}. To further boost the performance, we extend the existing detector by introducing contextual information using colormap (Section \ref{colormap}) for generating feature maps and prior group distribution for box refinement (Section \ref{pgd}). To evaluate the performance of our UISCGD, we propose our UI screenshot dataset (Section \ref{dataset1}). 
To summarize, we make the following contributions:
\begin{itemize}
\item A novel UI element grouping method inspired by previous perceptual group and navigation group, which contributes to multiple downstream UI-related software engineering tasks.
\item A robust, data-driven semantic component groups detector based on enhanced Deformable-DETR with a fusion strategy on color representation and a learned prior group distribution specialized for mobile GUIs, which achieves high performance. 
\item An empirical study on how our semantic component groups optimize perceptual group generation, improve code structure for UI-to-code automation, and generate accessibility metadata for screen readers.
\end{itemize}
\section{Background and Related Work}
\label{sec:Background}
\subsection{UI Element Detection}
To prevent any unnecessary confusion, we first draw a distinction between UI element detection and UI element grouping. UI element detection, as a specialized application of object detection, aims to derive a collection of bounding boxes from a image, each associated with a specific class label. When such detection methods are tailored for user UI screens, the resulting output corresponds to an assortment of UI elements present on the screen. Generally, the target object on GUI can be further summarized as text and non-text elements \cite{uied}. For text detection, OCR tools show high performance in UI tasks.  Earlier work like Tesseract \cite{23}, which adopts classic image binary maps to find characters by localizing the outline, shows significant effectiveness on UI screens which clear structural arrangement. Subsequent work \cite{24,10185179,10176817} has delved into the challenges of rotation, distortion, warping, and blurring that can occur in more complex real-world environments, significantly enhancing the robustness of detection algorithms. When applied to the relatively simpler scenarios in UI, these methods yield excellent results. 

Moving onto the category of Non-text elements, various methodologies have been deployed to recognize and categorize these elements within the UI space. Some work \cite{uied,perceptual_group} does not differentiate the types of Non-text elements, but faithfully identifies all eligible element objects, including but not limited to icons, widgets, and images. These methods typically do not rely on supervised learning, but utilize region-based segmentation algorithms \cite{SUZUKI198532,Torbert2016} to identify elements. Generally, these methods can serve as a good baseline for further research and applications. While other works \cite{10.1145/3491102.3502042,10.1145/3491102.3502073,vins,icon_annotation} focus on specific categories of non-text elements. Generally, elements are defined into dozens or even hundreds of categories based on differences in function and appearance. These methods often require a sufficiently large labeled dataset \cite{rico,10.1145/3544548.3581158,Screen_Recognition} for support. Like OCR methods, general object detection techniques applied in animals, human beings or vehicles have also been widely repurposed for use in the UI domain. Earlier works \cite{25,26} adopt traditional feature-based computer vision algorithms (e.g., Canny and SIFT) to detect UI widgets. Fully taking advantage of deep learning, some recent works \cite{27,28} borrow the state of the art (SOTA) model (like YOLO) of generic detection and achieve good performance. To achieve better performance in the UI domain, some researchers have also attempted to incorporate UI-specific features into the model training process. For example, some works \cite{UILM,vins,xiao2022ui} utilize additional modality information and others \cite{34} gather prior knowledge from large-scale data before training. 

\subsection{UI Element Grouping}
The grouping of UI elements, refers to the process of combining multiple UI elements together. This concept is often applied in situations where different UI elements are functionally related, or aesthetically grouped together to form a composite design element in the UI screen, as presented in Fig. \ref{fig1}. Although the grouping of UI elements can enhance the understanding of the overall UI layout and the interaction between its components, it generally appears as an intermediate process in specific downstream tasks \cite{10.1145/3472749.3474763}. Most grouping tasks \cite{uied, perceptual_group,Screen_Recognition,10.1145/3472749.3474763} involve or are based on a first-pass step, such as obtaining position information of elements from UI elements detection. Compared to the process of UI elements detection which always involves specific techniques and unique considerations, the formation of groups has not been sufficiently focused on. It usually relies on heuristic or hand-crafted rules established based on the needs of downstream tasks. For instance, the work in \cite{perceptual_group} uses Gestalt laws \cite{45} to achieve grouping. Specifically, they manually adjusted the relevant parameters and used clustering methods to achieve connectedness, similarity, proximity, and continuity among elements. For Screen Recognition \cite{Screen_Recognition}, however, the grouping goal is achieved by relying on the alignment characteristics and distance influence among UI elements. These varying rules make it difficult for one grouping method to be easily transferred to other tasks. In this paper, we aim to explore a more widely applicable form of UI element grouping, which would facilitate a deeper understanding of UI structures and contribute to the efficiency of downstream UI tasks. 

We believe that the Semantic Component Group proposed serves as a promising attempt. Structurally, it has excellent capabilities for both upward extension to section-level groups and downward decomposition to single elements, allowing for efficient utilization in tasks involving UI structure understanding \cite{perceptual_group, UILM} with an acceptable additional cost. Moreover, the elements within each group form semantic complements to each other, making this model highly adaptable to tasks involving semantic understanding such as UI summarizing \cite{10.1145/3472749.3474765} and those dealing with incomplete accessibility data \cite{Screen_Recognition}. In terms of the grouping methodology, we abandon the two-step approach used in previous works, focusing all our effort on the grouping step instead. Specifically, we no longer consider individual UI elements as distinct objects. Instead, we created a new semantic component group dataset, in which each group is treated as a standalone object. In our training for group detection, we incorporated additional unsupervised region segmentation algorithms \cite{uied} to generate feature maps about the independent position information of text and non-text elements. Furthermore, we utilized a structural prior learned from our dataset to guide the regression of group prediction errors.

\subsection{UI Elements Grouping for Better Code Generation}
Systematic GUI development usually engages a large team of designers and engineers \cite{29}. As the prototype designers and front-end code engineers work interdependently, any revision leads to repeated adjustments and takes much time. To tackle the repetitive aspects of GUI development, previous works \cite{30,31,32} adopted automatic tools in GUI code implementation. Given the GUI design prototypes created by design tools (e.g., Figma, Sketch, and Photoshop), automation tools perform a UI-to-code generation process. Technically, many implementations \cite{30,31} adopt a specific layout structure decomposition algorithm centered with a UI element detection. Great convenience has been observed by introducing deep learning technologies to this process, such as utilizing the vision method to extract UI elements from the background. Although the SOTA object detection methods achieve pixel-level accuracy for UI element detection, they are usually unaware of the UI structures, which causes problems with the generated client-side code, such as a highly redundant code structure for UI components in similar appearance \cite{perceptual_group}. In addition, current automation tools are unaware of the transformation of element representation during the manual process. For example, the ``watch'' icon we present in Fig. \ref{fig1}(a) consists of three basic UI elements in prototypes. While front-end engineers only deliver it as a single image in the code according to the design specification. Chen et al. \cite{UILM} reported this issue and observed significant code improvement with their element grouping. In this paper, by using our semantic component groups, we aim to solve another problem of structure loss, i.e., the organization and hierarchical information of UI elements in prototypes are not inherited in the generated code.

\section{Semantic Component Group Detection}
In this section, we introduce our method for semantic component group detection. We start by introducing our UI semantic component group dataset, as a pixel-only approach, we do not rely on metadata from design prototypes but UI screenshots and annotations for groups. To obtain groups, we then elaborate on the use of the state-of-art Two-Stage Deformable-DETR \cite{19} with iterative bounding box refinement. In a bid to further enhance performance, we introduce strategies involving the use of a colormap and prior group distribution.

\subsection{UI Semantic Component Group Datatset}
\label{dataset1}
Existing research has already accumulated a plethora of large-scale UI datasets, such as Rico \cite{rico}, AMP \cite{Screen_Recognition}, CLAY \cite{10.1145/3491102.3502042}, and WebUI \cite{10.1145/3544548.3581158}. However, most of these datasets only provide overall data information about UI pages or annotations for the positions and types of individual UI elements, offering little assistance for the detection of our UI semantic component groups. Additionally, UI iterations occur at a rapid pace, making the utilization of the latest real-world UI data beneficial. Driven by these two reasons, we proposed a dataset aimed at identifying UI semantic component groups. Statistically, we obtained 1989 mobile GUI screenshots from more than 200 Sketch/Figma prototypes, encompassing a variety of real-world UI apps in categories such as finance, shopping, music, and travel. After we exported all prototypes as UI screenshots, 10 workers were hired to annotate the semantic component groups in these screenshots. During the annotation, the workers determined a bounding box for each semantic component group in the form of $[x,y,w,h]$. To ensure the dataset accurately reflects the structure and semantics of UI component groups, we established specific annotation guidelines for our workers. First, the bounding boxes should strictly encompass both image and text parts. Second, the annotations must closely hug the edges of the elements, ensuring no extraneous elements are included. Third, if the aspect ratio of a bounding box is less than 1:8 or greater than 8:1, it should be adjusted to fall within this range to ensures that the bounding boxes are not excessively narrow or wide. To ensure quality and correct operational errors, one researcher randomly checked 20\% of the screens after the initial annotation and summarized errors for re-annotation. The workers then repeated the annotation based on our findings. Another round of checks and re-annotations was held to achieve our final data. As a result, we obtained 15167 semantic component groups in total. Table \ref{tab:dataset1} shows the
statistics of our semantic component group dataset. Our detector is trained with the training set. 

\begin{table}[htp]
\centering
\caption{Dataset statistics}
\vspace{0.15in}
\label{tab:dataset1}
    \begin{tabular*}{0.7\linewidth}{@{\extracolsep{\fill}}l c c }
    \toprule
    Split & Screenshots & Group Objects \\
    \midrule
    Training & 1591 & 11937 \\
    Validation & 199 & 1499 \\
    Test & 199 & 1731 \\
   \midrule
 Total & 1989 & 15167  \\
    \bottomrule
  \end{tabular*}
\end{table}

\subsection{Two-Stage Deformable-DETR with Iterative Bounding Box Refinement}
Combining multi-scale convolution neural networks and Transformer encoder-decoders, the Deformable-DETR has been considered a powerful object detector with both simple architecture and competitive performance. To decide our baseline detector, we see how it outperforms other SOTA architectures, as shown in Table \ref{tab:semantic group detection}. We briefly review its key framework to elucidate how the prior group distribution strategy is applied. Readers can refer to \cite{19} for a more detailed explanation.

\begin{figure}[tbh]
    \centering
    \includegraphics[width=0.8\linewidth]{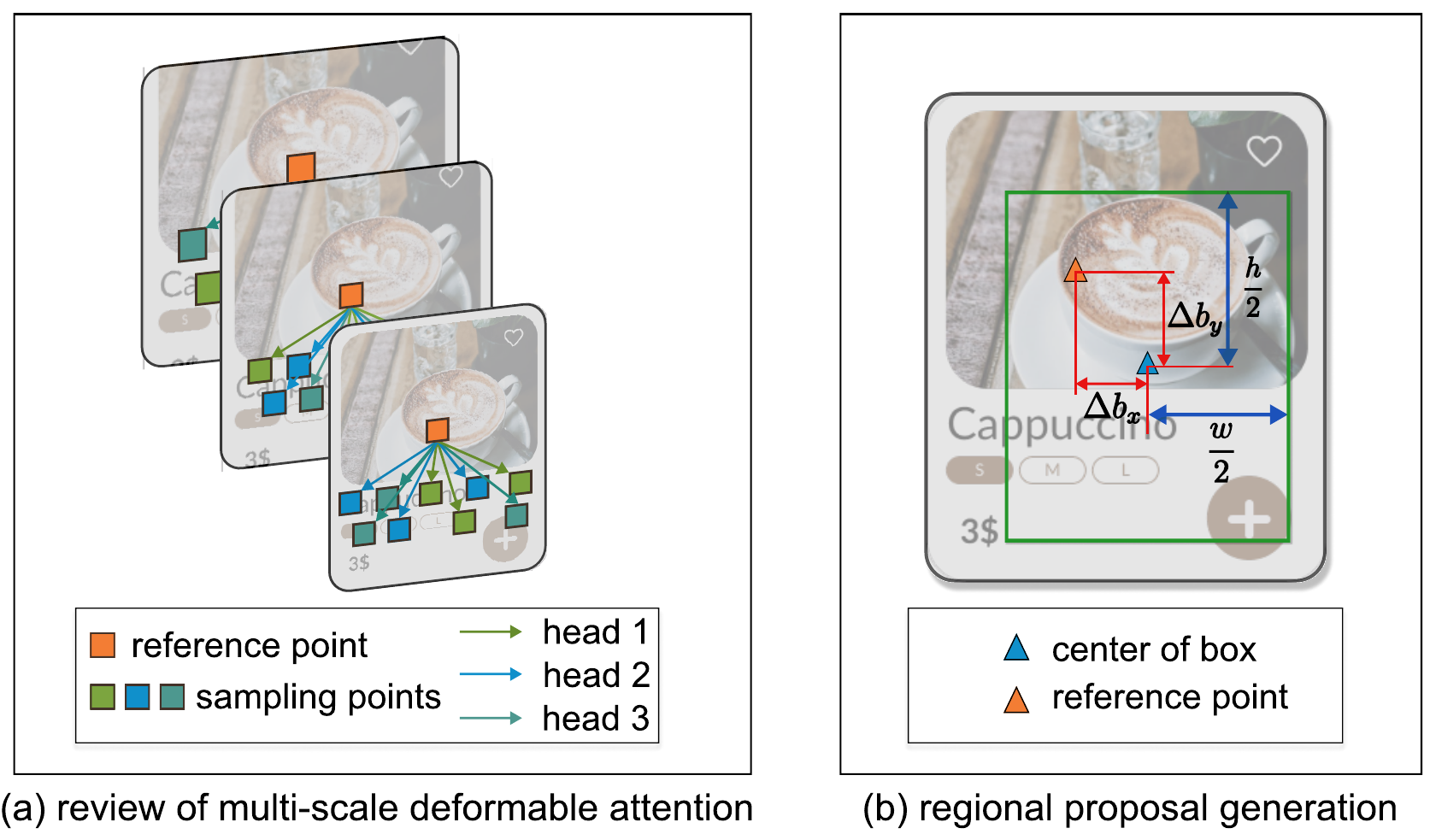}
    \caption{(a) A review of multi-scale deformable attention, attention mechanism apply between each reference point (in orange) and several points sampled; (b) the predicted bounding boxes represented by center point and box size is refined iteratively.}
    \vspace{-0.1in}
    \label{fig2}

\end{figure}

We first explain the multi-scale deformable attention module as 
\begin{equation}
\scalebox{0.85}{
$\begin{aligned}
   MSDeformAttn(z_q,\hat{p}_q,\{x^l\}_{l=1}^L) = \Sigma_{m=1}^M W_m [\Sigma_{l=1}^L \Sigma_{k=1}^K A_{mlqk} W_m^{'} x^l(\phi(\hat{p}_q)+\Delta p_{mlqk})].
\end{aligned}$
}
\end{equation} 
Fig. \ref{fig2}(a) shows how it is processed. It made two contributions from the original feature map attention \cite{33}. First, it utilizes $L$ input feature maps from different scales to capture objects that vary in size. Second, given each reference point (also known as query point) marked as orange, only $K$ points are sampled for calculation instead of all pixels. For $M$ attention head applied, each reference point gets $MK$ sampling points in total. The two-dimensional reference point $\hat{p}_q$ is normalized into [0,1] to unify its position in different feature maps. The $\phi$ function is used to re-scale it back to the original coordinates. $A$ denotes the attention weights and is normalized by $\Sigma_{l=1}^L\Sigma_{k=1}^K A_{mlqk} = 1$. And the sampling points are obtained by adding the offset $\Delta p_{mlqk}$. 

The term ``Two-Stage'' denotes the way to acquire the prediction bounding boxes. In the first stage, the model predicts a set of bounding boxes as the region proposals by 
\begin{equation}
\scalebox{0.85}{
$\begin{aligned}
   \hat{b}_i = \{\sigma(\Delta b_{ix}+\sigma^{-1}(\hat{p}_{ix}), \sigma(\Delta b_{iy}+\sigma^{-1}(\hat{p}_{iy}), \sigma(\Delta b_{iw}+\sigma^{-1}(2^{l_i-1}s), \sigma(\Delta b_{ih}+\sigma^{-1}(2^{l_i-1}s) \},
\end{aligned}$
}
\end{equation} 
where ${\Delta b_i\{x,y,w,h\}}$ is obtained by feeding the output feature maps of the encoder into an FFN regression head, ${p_i\{x,y\}}$ is the set of reference points, $\sigma$ denotes the Sigmoid function, and $s$ is set to 0.05. As shown in Fig. \ref{fig2}(b), for each initial reference point labeled as orange triangle, the center of regional proposal as blue triangle is obtained by adding ${\Delta b_i\{x,y\}}$. And its size $(w,h)$ is determined by  ${\Delta b_i\{w,h\}}$.
In the second stage, iterative bounding box refinement is applied at every decoder layer. For a $D$ layers decoder, given ${b_q^{d-1}}$ is the bounding box predicted by the $(d-1)^{th}$ layer, the $d^{th}$ layer refines the box as 
\begin{equation}
\label{eq3}
\scalebox{0.82}{
$\begin{aligned}
   \hat{b_q^d} = \{ \sigma( \Delta b^d_{qx}+ \sigma^{-1}(\hat{b}^{d-1}_{qx})), \sigma( \Delta b^d_{qy}+ \sigma^{-1}(\hat{b}^{d-1}_{qy})), \sigma( \Delta b^d_{qw}+ \sigma^{-1}(\hat{b}^{d-1}_{qw})), \sigma( \Delta b^d_{qh}+ \sigma^{-1}(\hat{b}^{d-1}_{qh}))\},
\end{aligned}$
}
\end{equation} 
where $\Delta b^d_{q\{x,y,w,h\}}$ denote the box offset predicted by an FFN regression head at $d^{th}$ layer, and the initial box $\hat{b}^0_{q\{x,y,w,h\}}$ is set to $\{\hat{p}_{qx},\hat{p}_{qy},0.1,0.1\}$.

\begin{figure}[thb]
    \centering
    \includegraphics[width=0.98\linewidth]{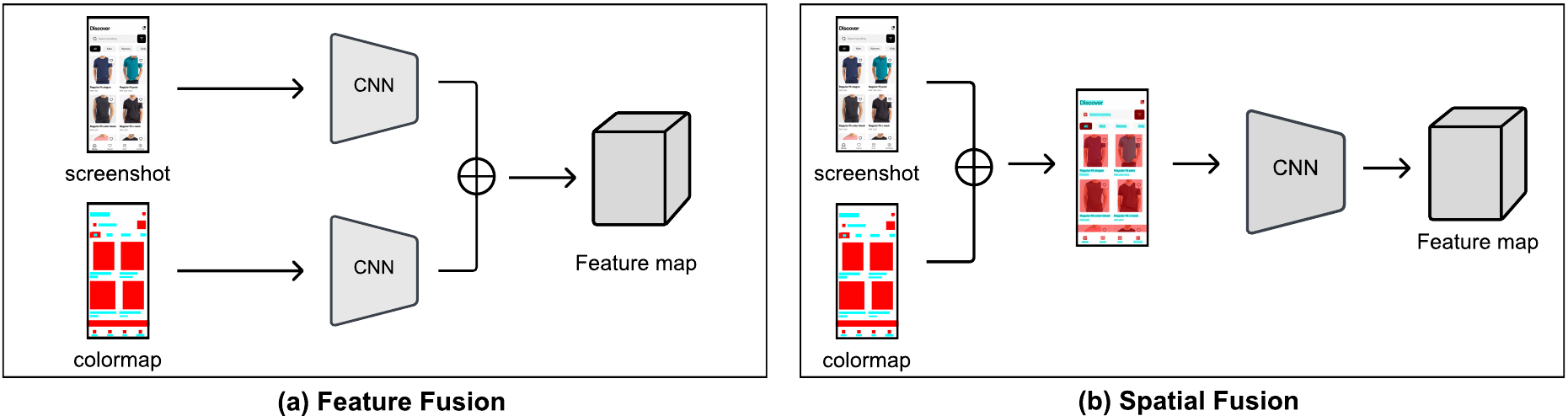}
    \caption{Two fusion strategies applied for colormap.}
    \vspace{-0.1in}
    \label{fig3}
  
\end{figure}

\subsection{UIED-based Colormap}
\label{colormap}
Previous works \cite{icon_annotation,vins} have addressed an inter-class variance of UI element detection, i.e., elements from the same class may vary in size and pixel representation. Although we do not separate our semantic component groups into multiple categories, we face a similar challenge. Our groups also vary greatly, especially in size and aspect ratio, as we present in Fig. \ref{fig1}(d). To enhance the detection performance, we encode a visual color representation of the view hierarchical structure of input images, called colormap. As a pixel-only method, we can not directly get information from the prototype metadata. Instead, we adopt UIED \cite{uied}, which performs unsupervised detection of text and non-text elements. Given a screenshot image, UIED detects the non-text elements with a combination of the flood-filling algorithm and Sklansky's algorithm, and we fill all these positions with red color. Then, the text elements are detected using the Google OCR tool, and we fill them with blue. The positions of text elements are filled after non-text elements, in which case texts inside images will also be recognized. Given a colormap shown in Fig. \ref{fig3}, we fuse it with the original image input so that prior knowledge is added that helps to assign sampling points. We tried two fusion strategies: feature fusion (a) and spatial fusion (b). For feature fusion, we obtain the deep-level feature of colormap and screenshot separately using a CNN feature extractor. Then the sum of the two feature maps is utilized for further box prediction. While for spatial fusion, we first make the superposition of the colormap and screenshot and feed it into the CNN. As for results, spatial fusion only boosts the detection performance with 0.4\% in precision, while feature fusion gives 2\% in precision, in which case we decide to use the feature fusion strategy. 

\begin{figure}[tbh]
    \centering
    \includegraphics[width=0.9\linewidth]{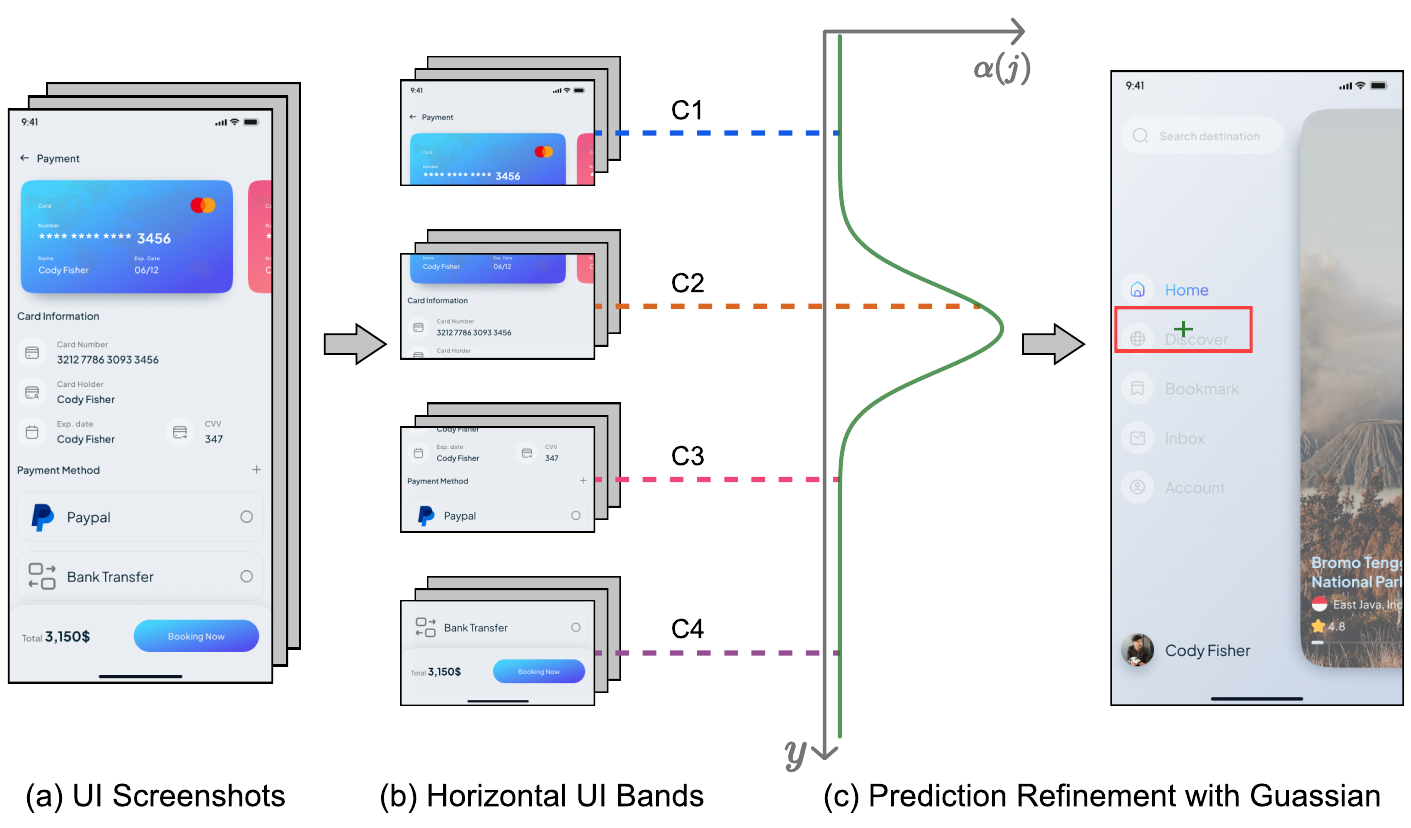}
    \caption{The Gaussian function $\alpha(i,j)$ applied a soft weighting for each local correlation on the box refinement. }
    \vspace{-0.1in}
    \label{fig4}
   
\end{figure}

\subsection{Prior Group Distribution}
\label{pgd}
In this section, we introduce our idea of exploiting spatial relationships of semantic component groups as prior learned knowledge to improve detection accuracy. The semantic component groups usually vary along the vertical direction while sharing a similar look in the same region. For example, the semantic component groups of the tab (icon and text), as shown in Fig. \ref{fig1}(d), usually appear at the bottom of the UI page and are of a similar small size. While picture groups (picture and subtitle) in Fig. \ref{fig1}(d) are more likely to be found in the middle. Additionally, similar groups often appear together, thereby forming UI sections with more comprehensive functionalities. The difference in size, aspect ratio, and position of such groups would have insight for box prediction. In this case, we estimate the correlation information of semantic component groups in the local region. 

As shown in Fig. \ref{fig4}, given screenshots from our training set, we divided each into $N$ horizontal bands to better capture the unique vertical distribution of semantic component groups in different UI regions. The varying layouts and positional characteristics of these groups across the bands reflect the typical organization of elements in a standard UI design. For group boxes in each band, we calculate the correlation matrix $C$ of [$x_{center}$,$y_{center}$, $width$, $height$]$\in \mathbb{R}^{4\times 4}$. Then we normalize it by row and column and make it into $[0,1]$. Inspired by \cite{34}, we add another box refinement similar to Equation \ref{eq3} based on group correlation, where $W$ are the weights of a 3-layer MLP including the bias, $\Delta b'^d_{q\{x,y,w,h\}}[i]$ of the $i^{th}$ prediction box is obtained as
\begin{equation}
\label{eq4}
\begin{aligned}
   \Delta b'^d_{q\{x,y,w,h\}}[i] = \sum_{j=1}^N \alpha(i,j)\cdot C b'^{d-1}_{q\{x,y,w,h\}}[i]W,
\end{aligned}
\end{equation} 
where $\alpha(i,j)$ denotes the Gaussian influence of $j^{th}$ correlation matrix on the $i^{th}$ prediction box. It is calculated as
\begin{equation}
\begin{aligned}
   \alpha(i,j) = \frac{1}{\sqrt{2\pi\sigma^2}}exp^{-\frac{1}{2}(\frac{d-\mu}{\sigma})^2},
\end{aligned}
\end{equation} 
where $d$ is the distance between the center of $i^{th}$ prediction box and $j^{th}$ band. This newly acquired bounding box offset $\Delta b'^d_{q\{x,y,w,h\}}$ related to prior group distribution is added into Equation \ref{eq3}, where it jointly contributes to overall box refinement. During inference, the prediction box is tuned based on the group distribution with this extra refinement. As an example in Fig. \ref{fig4}, with a $j^{th}$ predicted bounding box centered at (x,y) shown as the green cross marker, the Gaussian influence shown in green illustrates the distribution of $\alpha(j)$ with the vertical position of the bounding box $y$ as the independent variable. In this case, As the predicted bounding box is closer to second band in the vertical direction, we expect the correlation guided by $C_2$ to have a greater influence on it based on Equation \ref{eq4}.

\section{Empirical Evaluation}

\subsection{Accuracy of GUI Semantic Component Group Detection}
To evaluate the effectiveness of our semantic component group detection, we employ a three-step assessment. Firstly, we compare the accuracy of our method with other established baseline methods. Secondly, we conduct an ablation study to demonstrate the efficacy of each proposed strategy. Lastly, we provide a visualization of the attention process in group detection to better understand how our improvements are implemented.

\subsubsection{Baseline Methods}

As our implementation is rooted in deep visual methods, we compare our approach with prominent object detection techniques. RetinaNet \cite{37} represents an anchor-based approach that scans the feature map with pre-defined anchor boxes, while YOLOX \cite{38} exemplifies an anchor-free approach that assigns targets by predicting the object's center and size. We selected these two methods as baselines because our UISCGD borrows concepts from both: it assigns anchor boxes and fine-tunes their center and size via box refinement. We also choose Faster-RCNN \cite{fatser_rcnn} due to its similar region proposal strategy, and VarifocalNet \cite{40} for its comparable box refinement concept used in UISCGD. For all these methods, we follow the reported parameter settings in the original sources and start training from the initial state using our UI dataset. 

In addition to comparing supervised methods based on deep learning, we also consider unsupervised approaches. Notably, perceptual grouping \cite{perceptual_group} involves an intermediate stage at the component level and deals with containers that share a compositional similarity with our groups. Therefore, we incorporate it as one of our baseline models. As an unsupervised approach, we adopt the same parameter settings as used for mobile UI \cite{uied}. To gauge performance, we employ the average result of $precision$, $recall$, and $F1~score$ of IoU@[0.5:0.95] as metrics.

\subsubsection{Implementation Details}
ImageNet \cite{35} pre-trained ResNet-50 \cite{36}
was utilized as our backbone. For the feature fusion strategy, we adopted another backbone for colormaps. We followed the same setting for Transformer as \cite{19}, i.e., $M=8$, $K=4$, and $D=6$. For the prior group distribution strategy, we set hyperparameters $N=4$, $\sigma=0.3$, and $\mu = 0$. We trained our model with a mini-batch of 2 for 50 epochs using the SGD optimizer with a momentum update of 0.9 and a weight decay of 0.0001. The initial learning rate was set at 2.5e-5 and was decayed at the $40^{th}$ epoch by a factor of 0.1. The model was trained on an NVIDIA GeForce RTX 3090 GPU and took about 8 hours to converge.

\subsubsection{Results}
Table \ref{tab:semantic group detection} shows the overall performance of semantic component group detection. As a result, UISCGD achieves a much higher F1 score (0.775), which is 6.1\% higher than the second-best model (VarifocalNet). This indicates that compared to the supervised methods, UISCGD provides superior performance in semantic component group detection. In addition, UISCGD also achieves 23.5\% increase in F1 score with the container group, which is a UI-specified method and produce similar component-level groups. As a unsupervised approach, the container groups are identified strictly based on their rectangular shape and other geometric rule. However, the rigid, handcrafted requirements of the container group methodology could limit its ability to identify certain semantic component groups. Table \ref{tab:ablation study} shows the ablation of our UISCGD, where ``-'' means dropping the current component and PG denotes our prior group distribution strategy. As we can see, removing our colormap and prior group distribution strategies led to a significant drop in precision (0.058) and recall (0.046). Either colormap or prior group distribution boosts the performance, with a slight increase in F1 score: 2\% for colormap and 2.1\% for prior group distribution. The detection column in first row of Fig. \ref{fig6} presents the detection results of our UISCGD, where we draw the bounding boxes in green rectangles. To show the robustness of UISCGD, case 1 is randomly chosen from our UI dataset. Case 2 is obtained from \cite{perceptual_group} in Android platform. And case 3 is downloaded from the Figma community.

\begin{figure}[tbh]
    \centering
    \includegraphics[width=0.93\linewidth]{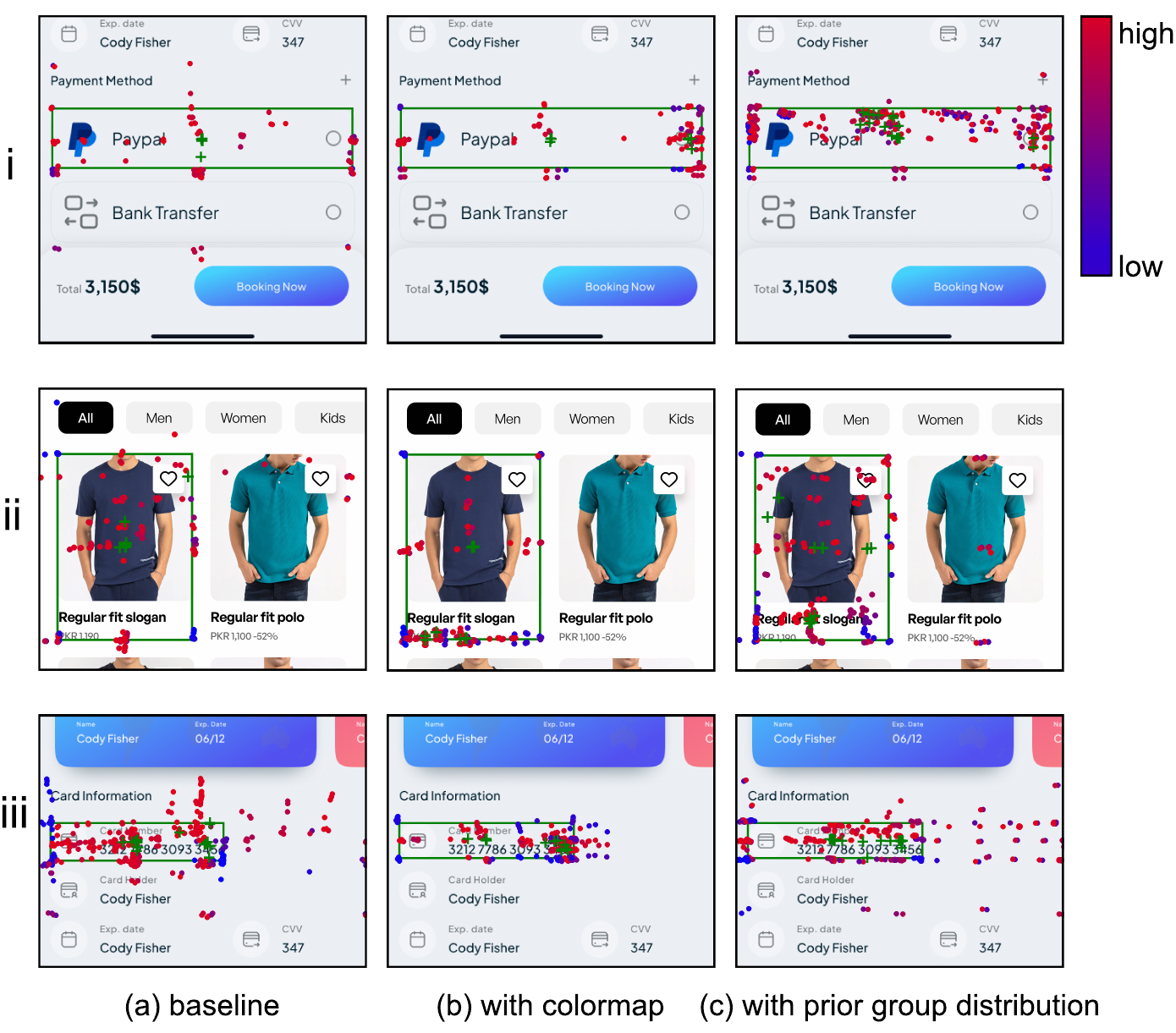}
    \caption{Visualization of the distribution of reference points and sampling points in muti-scale deformable attention (green box - ground truth, green cross marker - reference point, filled circle - sampling points).}
    \vspace{-0.1in}
    \label{fig5}
    
\end{figure}

To further discuss why our approaches boost the performance of semantic component groups detection, Fig. \ref{fig5} visualizes the multi-scale deformable attention of our UISCGD. The reference points, also known as query points, are shown as green cross markers. Each sampling point is marked as a filled circle whose color indicates its attention weight. We skip those
sampling points that are not apparent (with attention weight\textless0.1) for readability and label the
ground truth of groups in green bounding boxes. Comparing (a) and (b), we can see the spatial knowledge from our colormaps makes the sampling points focus closer on the group.  For example, in case 2, many sampling points in (a) look at pixels far from the group while in (b) they are restricted near the potential group, in which
case the attention results contain more information about the group itself and the adjacent context. (a) and (c) show the effect of our prior group distribution strategy. This strategy formulates the relationship between UI semantic component group shape (aspect ratio) and its location on the UI screen by adding a bias to the distribution of sampling points. For example, in case 3, the attention focuses more on pixels horizontally because the prior group distribution infers that groups in this area are more likely to have small aspect ratio. 
\begin{table}[tbh]
	\centering
	\caption{Performance Comparison: GUI Semantic Component Group Detection (IoU@0.5:0.95)}
        \vspace{0.15in}
        \begin{tabular*}{0.9\linewidth}{@{\extracolsep{\fill}}l c c c}
		\hline
    \textbf{Method} & \textbf{Precision} & \textbf{Recall} & \textbf{F1} \\
		\hline
            RetinaNet        & 0.550 & 0.702 & 0.617 \\
		Faster-RCNN        & 0.652 & 0.757 & 0.701 \\
		VarifocalNet       & 0.649 & 0.793 & 0.714 \\
		YOLOX       & 0.634 & 0.761 & 0.692 \\
            Container Group  &0.643  & 0.466 & 0.540  \\
		UISCGD     & \textbf{0.706} & \textbf{0.858} & \textbf{0.775} \\ 
	\hline
	\end{tabular*}
	\label{tab:semantic group detection}
\end{table}


\begin{table}[tbh]
	\centering
	\caption{Ablation Study: GUI Semantic Component Group Detection (IoU@0.5:0.95)}
        \vspace{0.15in}
        \begin{tabular*}{0.9\linewidth}{@{\extracolsep{\fill}}l c c c}
		\hline
    \textbf{Method} & \textbf{Precision} & \textbf{Recall} & \textbf{F1} \\
		\hline
            UISCGD     & \textbf{0.706} & \textbf{0.858} & \textbf{0.775}\\
		UISCGD-PG  & 0.668 & 0.832 & 0.741 \\
            UISCGD-colormap    & 0.665 & 0.839 & 0.742 \\
            UISCGD-colormap-PG & 0.648 & 0.812 & 0.721 \\
	\hline 
	\end{tabular*}
	\label{tab:ablation study}
\end{table}

\begin{figure*}[tbh]
    \centering
    \includegraphics[width=0.96\linewidth]{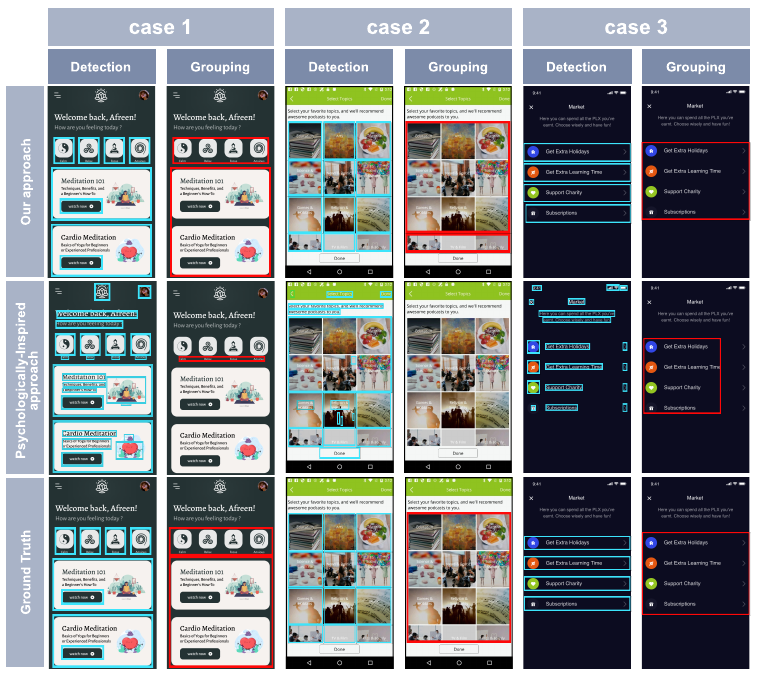}
    \vspace{-0.02in}
    \caption{Examples of semantic component group detection and perceptual grouping results. All detection results are labeled in blue. And perceptual groups derived are represented in red.}
    \label{fig6}
\end{figure*}

\subsection{Perceptual Group Performance}
\label{perceptual group result} 

Perceptual grouping, named after the Gestalt laws of perceptual organization \cite{45}, illustrates the phenomenon that the human mind tends to partition a set of physically discrete elements into groups. This cognitive process involves a series of grouping principles including connectedness, similarity, proximity, and continuity \cite{britannica2008britannica}. Despite the widespread use of Gestalt principles-based perceptual grouping in UI design \cite{koch2016computational,uxmisfit2019gestalt} and evaluation \cite{hovorushchenko2019method,macnamara2017evaluating} to validate structural rationality, research on automatically inferring perceptual groups from UI pages has been relatively sparse. Systematic studies on this topic can be traced back to the work of Xie et al. \cite{perceptual_group}. However, due to the nested nature of UI grouping structures (i.e., smaller groups combine to form larger ones), and since different requirements and technical tasks may necessitate various forms of groups, the concept of perceptual grouping is not definitive. To narrow down our discussion, in this paper, our concept of perceptual grouping follows the definition set forth by Xie et al. \cite{perceptual_group}, specifically referring to the section-level group as illustrated in Fig. \ref{fig1}(c). In UI pages, such section-level groups are typically presented in forms such as cards, lists, multi-tabs, and menus. Identifying these groups can help us determine which actions are suitable for specific parts of the GUI (clicking navigation tabs, expanding cards, scrolling lists), making automatic GUI testing more efficient \cite{li2019humanoid,element_interaction}. Moreover, through perceptual grouping, modular and reusable GUI code can be automatically generated from GUI design images, accelerating the rapid prototyping and evolution of GUIs \cite{25,29}. In this section, we first introduce our approach to generate perceptual groups of GUI based on our semantic component group detection. We then discuss the small prototype dataset that we established to evaluate the performance. The results we present show that our approach infers perceptual groups reliably.

\subsubsection{Method}
 For Xie et al.'s psychologically-inspired perceptual grouping \cite{perceptual_group}, their initial step involves pinpointing the locations of all single text and image elements \cite{uied}. Based on this, they sequentially employ the principle of connectedness to aggregate proximate text and image elements, yielding an effect akin to our semantic component groups. Subsequently, they utilize the principle of similarity to extend similar small groups into section-level groups, and finally, iterative grouping adjustments are made through the principle of proximity. As a contrast, our approach bypasses the processing of single text and image elements. Instead, by leveraging the principles of similarity and proximity, we assess whether semantic component groups belong to the same section-level perceptual group based on the similarity in size among these groups and their spatial proximity. We develop heuristics that merge our semantic component groups in Algorithm. \ref{algorithm_1}. Given a GUI screenshot and all its semantic component groups represented with bounding boxes in a quadruple notation  $[x,y,w,h]$ predicted by our UISCGD, the algorithm takes two steps to obtain all perceptual groups. 

We start by aggregating all semantic component groups based on size (i.e., width and height). As we allow some deviations between the predicted width and height of a semantic component group and the ground truth, for example, $IoU=0.90$, semantic component groups with similar appearance could be predicted into different shapes as shown in the detection column of case 3 in Fig. \ref{fig6}. To overcome inaccurate predictions, we adopt the Density-Based Spatial Clustering of Applications with Noise (DBSCAN) algorithm \cite{42} to implement the clustering, which is insensitive to those outliers and still performs qualified clustering. In addition, the DBSCAN requires no pre-defined cluster number $k$ like in the K-means algorithm, which satisfies our need that the number of perceptual groups varies for different GUI pages. The number of core points is set to $MinPts=2$, and to determine the value of eps-neighborhood, we visualize the curve of the K-distance function and take the inflection point $eps=0.0116$. As a result, it provides $C$ clusters, each containing at least two semantic component groups of similar size. We adopt $cluster_c$ to represent the list of groups in the $c^{th}$ cluster.

We then form perceptual groups within each cluster based on their spatial relationship. In preparation, we define two linked lists: the output perceptual group list $O=\{\}$ and the remaining list $R=cluster_c$. Initially, we randomly select one element $bb_m$ in the $R$ and move it to the output perceptual group list $O$. For each pair of $bb_m$ in and $bb_n$ in the two linked lists, we focus on whether they are aligned. If they are aligned (either vertically or horizontally) and the minimum distance between two bounding boxes does not exceed our pre-defined connectivity threshold, we move $bb_n$ into the output perceptual group list $O$. Every time a new element is added to $O$, we repeat the pairwise checking procedure and finish the checking when no new element can be added to the output perceptual group list. Finally, if more than two semantic component groups ${[x,y,w,h]_{i=1}^{N}}$ are in the output list, the specification of corresponding perceptual group is determined as $[min(x),min(y),max(x+w),max(y+h)]$ in the form of the top left and bottom right points. For the legacy elements in $cluster_k$, we iteratively perform the above operations until no element is left. 
\begin{algorithm}\small
\caption{Perceptual Group Algorithm} 
\label{algorithm_1}
\begin{flushleft} 
\begin{tabular}{rl}
    \textbf{Input:} & Semantic component groups detected represented as $\{ bb_i \}_{i=1}^{N}$.\\
    \textbf{Output:} & Perceptual groups specified as $Res= \{ x_1,y_1,x_2,y_2 \}_{i=1}^{M}$.\\
    1: &  Aggregate semantic component groups with similar size;\\
    2: & $\{ bb_i \}_{i=1}^{N} \in [0,1] \gets$ bounding boxes normalization;\\
    3: & Initialize DBSCAN cluster $DBSCAN(eps=0.0116, MinPts=2)$;\\
    4: & Form $C$ clusters with $DBSCAN.fit(data)$;\\
    5: & Form perceptual group inside each cluster $cluster_c$;\\
    6: & \textbf{for all} $cluster_c$ in $Cluster$ \textbf{do}\\
    7: & \quad \textbf{while} $len(cluster_c) > 1$ \textbf{do}\\
    8: & \quad \quad Output list $O \gets {bb}_m$;\\
    9: & \quad \quad Remaining list $R \gets cluster_k-bb_m$;\\
    10: & \quad \quad \textbf{for all} $bb_m$ in $O$ \textbf{do}\\
    11: & \quad \quad \quad \textbf{for all} $bb_n$ in $R$ \textbf{do}\\
    12: & \quad \quad \quad \quad \textbf{if} Aligned($bb_m$, $bb_n$) \textbf{and} MinDist($bb_m$, $bb_n$) $< T$ \textbf{then}\\
    13: & \quad \quad \quad \quad \quad $O \gets O + bb_n$;\\
    14: & \quad \quad \quad \quad \quad $R \gets R - bb_n$;\\
    15: & \quad \quad \quad \quad \textbf{end if}\\
    16: & \quad \quad \quad \textbf{end for}\\
    17: & \quad \quad \textbf{end for}\\
    18: & \quad \quad $Res \gets Res + \{min(x), min(y), max(x+w), max(y+h)\}$ for $\{x,y,w,h\}$ in $O$;\\
    19: & \quad \quad $O \gets \{\}$;\\
    20: & \quad \textbf{end while}\\
    21: & \textbf{end for}\\
    22: & \textbf{return} $Res$.\\
\end{tabular}
\vspace{-4mm} 
\end{flushleft}
\end{algorithm}


\subsubsection{Prototype Dataset}
\label{prototype dataset}
To evaluate how close our predicted perceptual groups are to the actual designs, we collected 30 prototypes from the Figma community \cite{figma}. All these prototypes are created by experienced UI designers and have at least received 500 downloads (with the most popular one having more than 10k downloads). Each prototype contains several UI pages of a particular app. In contrast to the UI semantic component group dataset, which is constructed based on UI screenshots, the perceptual group dataset is derived from design prototypes. This means we can directly extract the positional information of the section-level perceptual groups (cards, lists, multi-tabs, and menus) from the view hierarchy parameters without the necessity for manual annotation. Statistically, we obtain 274 different UI pages and 631 perceptual groups.   

\begin{table}[tbh]
	\centering
	\caption{Performance: GUI Perceptual Group Detection (IoU@0.5:0.95)}
	\begin{tabular*}{0.9\linewidth}{@{\extracolsep{\fill}}l c c c}
		\hline
    \textbf{Method} & \textbf{Precision} & \textbf{Recall} & \textbf{F1} \\
		\hline
            Our approach        & \textbf{0.863} & \textbf{0.899} & \textbf{0.881} \\
		Psychologically-Inspired approach \cite{perceptual_group}        & 0.824 & 0.773 & 0.798 \\
		\hline
	\end{tabular*}
	\label{tab:perceptual group detection}
\end{table}

\subsubsection{Result}
To report the performance, we employ the same metrics used for semantic component group detection. We compare our approach with the psychologically-inspired grouping method proposed by Xie et al. \cite{perceptual_group}. The results in Table \ref{tab:perceptual group detection} show that our approach achieves a much higher F1 score (8.3\%) compared with the psychologically-inspired approach. The bottleneck of the psychologically-inspired approach appears mainly in recall which is 12.6\% worse than ours, which infers that our approach is more efficient in retrieving positive perceptual group samples. As an unsupervised method, the psychologically-inspired approach relies on several hand-craft parameters to perform elements merging. These parameters are tuned to achieve the best performance based on their Android app GUIs. While a cross-platform performance decay has been observed when testing on our prototype dataset, which also contains samples from Apple devices. Fig .\ref{fig6} case 1 demonstrates this issue, where the text and non-text elements detection still works well and offers convinced results in the second row of the first column. We use red boxes to denote text elements and green boxes for non-text elements. However, no perceptual group is detected from the grouping result in the second row of the second column because of the failure of the proximity check. For our approach, UISCGD is trained with dataset contains cross-platform samples which makes it performs accurate detection. Based on this, we can always recognize perceptual groups by setting a distance threshold with large tolerance in Algorithm \ref{algorithm_1}. The other two examples in Fig. \ref{fig6} also present our advantages. Case 2 comes from the Android dataset. We present this example to show that our approach performs better on GUIs with poor contrast ratio and indistinct boundaries between UI elements. The detection result in the second row of the third column shows that the psychologically-inspired approach fails in the early stage of element detection. The six pictures in the upper two rows of UI page are detected as single because of the fuzzy boundaries. In addition, most text lines inside pictures are missed as it is hard to recognize white text from a light background image which we refer to as the poor contrast ratio issue. As a result, no perceptual group is detected. The first row of the third column shows our detection of semantic component groups, which recognize all boundaries accurately. The first row of the forth column shows our grouping results based on Algorithm \ref{algorithm_1}. Because the three semantic component groups in the bottom of UI page are incomplete and are not the same size as the upper ones, we obtain two perceptual groups by our approach. In contrast, the ground truth shown in the last row of the forth column indicates that they should be considered a single perceptual group. As the page scrolls up and down, there are always semantic component groups that are displayed incompletely, while the display area is always the same. Case 3 comes from our UI screenshot dataset. Both approaches recognize the target perceptual group on this page, while our approach achieves a much higher IoU (0.905) with the ground truth than the psychologically-inspired approach (0.674). For the psychologically-inspired approach, it fails to merge the ``$<$'' icon in its connectedness test.

\subsection{Code Structure Improvement}
\label{code quality}
The view hierarchy of a UI prototype denotes how the UI elements are organized in design time. It reflects how designers place UI elements (i.e., pictures, text, and basic shapes) to form components, modules, and the whole page. Unfortunately, the structure of UI elements changes a lot when they are generated as code by automation tools. Xie et al. \cite{perceptual_group} discussed some common problems, such as code redundancy and structure loss. Generally, the code generated by automation tools is different from what experienced modular GUI code developers would write, in which case developers still need to do a lot of modification on the code. The poor practicality and usability undermine the purpose of automation to accelerate GUI development and make the life of developers easier. In this section, we present our semantic component group detection as an application that fills this gap in UI code automation.

\subsubsection{Environment Specification}
There are quite a few automation tools for UI code generation, while most of them remain as research demos or GitHub projects. To show the validity of our approach in actual production, We choose Imgcook \cite{22} as the automation tool. Imgcook is a mature commercial automation tool proposed by the front-end team of Alibaba and serves dozens of mobile apps for finance, travel, shopping, and other scenarios. People can use it through its web application, CLI, or design tool plug-in. In this experiment, we adopt Figma for prototype visualization and modification and combine the web app with the Figma plug-in of Imgcook for code generation and structure visualization. 

\begin{figure}[tbh]
    \centering
    \includegraphics[width=0.9\linewidth]{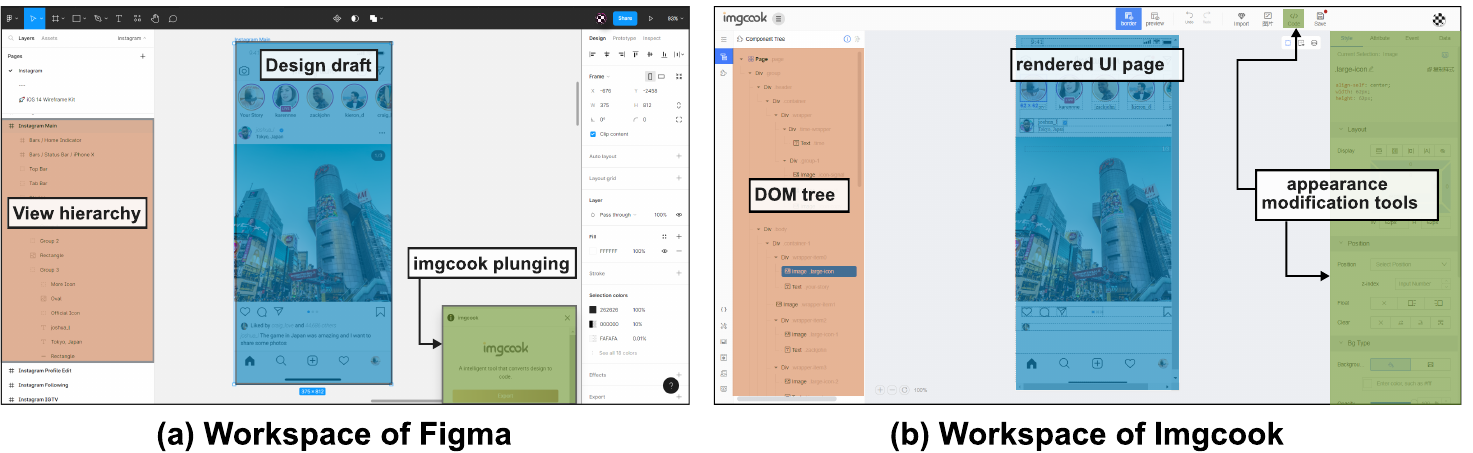}
    \caption{Environment used for code improvement experiment. (a) presents the Figma workspace for prototype design; (b) presents the Imgcook workspace for code generation.}
    \vspace{-0.1in}
    \label{fig7}
 
\end{figure}

An example of the application environment is shown in Fig. \ref{fig7}. Part (a) shows the workspace of Figma. The view hierarchy of the ``Instagram Main'' page selected is shown in the left part, which is a standard tree structure taking the container ``Instagram Main'' as the root node. Every non-leaf nodes in this tree represent a container holding several UI elements (as leaf nodes). By clicking the export button on the Imgcook plug-in, users will be directed to the code generator shown as (b). The left part shows a component tree (or DOM tree)  which visualizes the elements in the generated HTML as a tree structure. The center part displays the rendered UI page. Note that sometimes the elements may be missing or shifting, in which case the right part offers tools for appearance modification. After correcting all errors, users can generate the client-side code by clicking the ``Code'' button in the top toolbar. Multiple code architectures (e.g., Html5, React, Vue) and styles (JXS and TSX) are offered based on user preference. For readability, we compare the improvement in the follow-up experiment based on the DOM tree instead of a specific codebase.

\begin{figure*}[tbh]
    \centering
    \includegraphics[width=0.94\linewidth]{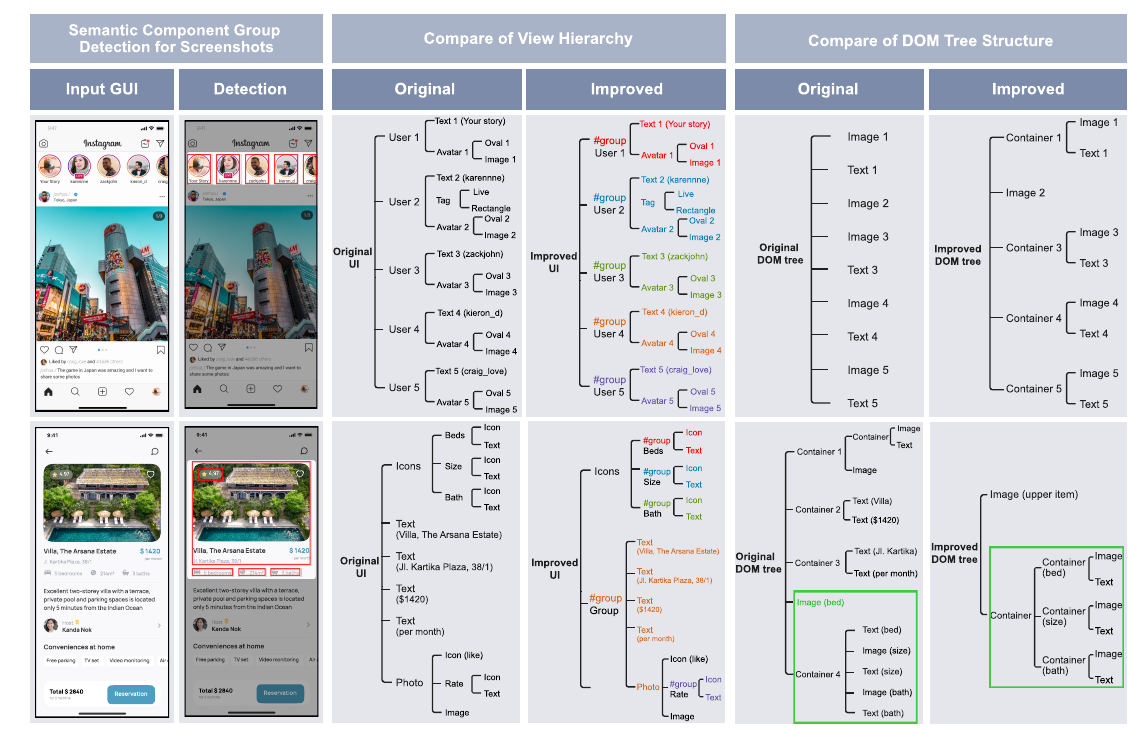}
    \caption{Examples of code structure improvement by utilizing semantic component groups. The first column displays the input GUI screenshots, and the second column presents the group predictions in the non-shadow regions using UISCGD. Columns three and four demonstrate modifications to the design prototype's view hierarchy using Algorithm \ref{algorithm_merge}, where we introduce additional markers indicating groups. The fifth and sixth columns show the structure of the generated code DOM Tree.}
    \label{fig8}
    
\end{figure*}

\subsubsection{Method}
Given a UI prototype, we export its preview as a screenshot and infer all semantic component groups with our UISCGD. The application that improves code structure is defined as a two-stage task: UI layer retrieval and code generation. As the bounding boxes predicted on the screenshot do not fully match the UI layers in the prototype, we apply a layer retrieve algorithm based on \cite{UILM} to find all the layers in prototype. The algorithm is shown as Algorithm~ \ref{algorithm_merge}. Given the predicted bounding boxes $\{ bb_i \}_{i=1}^{N}$ and view hierarchy of prototype $V_{tree}$, we first traverse $V_{tree}$ to get the list of all leaf nodes $\{ le_j \}_{j=1}^{M}$ which represent basic UI elements. For each bounding box, we traverse $\{ le_j \}_{j=1}^{M}$ and calculate the intersection area between the bounding box $bb_i$ and each $le_j$. For $le_j$ with an intersection area that exceeds the intersection threshold $T_i$, we save it to the list $temp$. For each layer node in $temp$, we calculate its depth difference with $T_m$, where $T_m$ is the majority value of depth in $temp$. The depth of a node in $V_{tree}$ is defined as the  shortest path length from the root. If the difference exceeds the distance threshold $T_d$, we remove it from $temp$. The filtered $temp$ is then saved to $\{ layer_i \}_{i=1}^{N}$ where $layer_i$ denotes the retrieved UI layers for $bb_i$.

Based on the retrieved layers for each semantic component group, we process the UI prototype and use it for code generation. Imgcook provides a group protocol for any non-leaf node in the UI prototype, ensuring every UI element in the current container will be constrained in the same DOM Tree node. In this way, we can form a rational DOM tree structure and further a usable code. To apply the prototype, the only thing to do is to name the target layer in prototypes with the prefix ``\#group\#''. Given the retrieved layers for a semantic component group, if a node in the view hierarchy contains exactly these layers (without any other elements involved), we apply the protocol on this node. Otherwise, a new node should be created.

\begin{algorithm}\small
\centering
\caption{Layer Retrieval Algorithm} 
\label{algorithm_merge}
\begin{flushleft}
\begin{tabular}{rl}
    \textbf{Input:} & $N$ predicted bounding boxes $\{ bb_i \}_{i=1}^{N}$ and view hierarchy of prototype $V_{tree}$.\\
    \textbf{Output:} & a list of UI layer groups $Res = \{ layer_i \}_{i=1}^{N}$.\\
    1: & $T_i \gets \text{pre-determined threshold of the intersection}$;\\
    2: & $\{ le_j \}_{j=1}^{M} \gets$ Traverse $V_{tree}$ to get leaf nodes for all UI layers;\\
    3: & \textbf{for} \textbf{all} $bb_i$ in $bb$ \textbf{do}\\
    4: & \quad \textbf{for} \textbf{all} $le_{j}$ in $le$ \textbf{do}\\
    5: & \quad \quad \textbf{if} $le_{j} \cap bb_i > T_i$ \textbf{then}\\
    6: & \quad \quad \quad save the layer $le_{j}$ to the list $temp$;\\
    7: & \quad \quad \textbf{end if}\\
    8: & \quad \textbf{end for}\\
    9: & \quad Filter $temp$;\\
    10: & \quad $T_d \gets \text{pre-determined threshold of the node distance}$;\\
    11: & \quad $T_m \gets \text{majority value of node depth in}~temp$;\\
    12: & \quad \textbf{for} \textbf{all} $l_{i}$ in $temp$ \textbf{do}\\
    13: & \quad \quad \textbf{if} $|depth(l_i)-T_m|>T_d$ \textbf{then}\\
    14: & \quad \quad \quad remove $l_i$ form $temp$;\\
    15: & \quad \quad \textbf{end if}\\
    16: & \quad \textbf{end for}\\
    17: & \quad remove the layers in $res$ from flatten list $fl$ and update $fl$;\\
    18: & \quad $Res \gets Res+temp$;\\
    19: & \textbf{end for}\\
    20: & \textbf{return} $Res$.\\
\end{tabular}
\vspace{-4mm}
\end{flushleft}
\end{algorithm}

\subsubsection{Result}

To show the improvement of code structure after applying our semantic component groups, we use the same prototype dataset described in Section \ref{prototype dataset}. Fig. \ref{fig8} shows two examples that we choose to illustrate the flaws in original UI code automation and how our semantic component groups improve them. The first column shows the input screenshots, and the second row shows the semantic component groups detected by our UISCGD. The third row reflects the original view hierarchy in the prototype. The fourth row is obtained after applying our layer retrieval algorithm, where all retrieved layers in the same semantic component group are labeled with the same color with the prefix ``\#group\#''  on their parent node. The fifth and sixth rows compare the differences in the elements DOM tree, which reflects the structure of code generated by Imgcook. To make it easier to understand, we only present the detection results, view hierarchy, and DOM tree structure of the problem area, highlighted in the second row.

The first example is related to the group with text and non-text elements arranged vertically. The displayed component includes five semantic component groups with a similar structure shown in the view hierarchy. Each group contains a text description and an avatar formed by an oval shape and profile image. An extra ``live'' tag is added on the second user profile which is slight different from others. Comparing the original and improved view hierarchy, we do not create any new containers or change any element positions. The two view hierarchies are highly similar except for these ``\#group\#'' markers on each user node. While comparing the two DOM trees generated by Imgcook, we can see that the original version loses almost all structural information. It is flattened with all text and images in the same container, and is hard to identify the actual groups. When we come to the code level, we will obtain a large bunch of repetitive code, which no UI code developer would accept. Moreover, it is also hard for code migration if we want to reuse one of the groups in another UI design. Developers need to figure out the exact code snippet. While in our improved version, the tree structure is kept with each container holding one group. Developers are allowed to make any changes more convenient as well as code migration. 

The second example contains two types of our groups. The detection results show that its top part is a more complex picture group with multiple texts in different styles and positions. Then the remained part in the bottom are three icon groups with elements arranged horizontally. Comparing the two view hierarchy trees, we can see a new container is created for the picture group with text elements describing the name, address, and price. The design is organized into two partitions by modifying the structure in prototypes. Designers can easily reuse or migrate any part on other designs by copying and pasting the whole subtree in the view hierarchy. For the generated code structure, it is recognized as four bands based on horizontal cutting in the original DOM tree. The elements in the green rectangle show how the three icon groups are organized in the generated code. Two problems arose in this area: first, the image ``bed'' labeled in green which represents one of the icon images should be in the same container with the text ``bed''. Second, the elements inside the container face the same issue as we discussed in the first example, where no group information remains. In contrast, our improved version solves both problems.

To quantitatively demonstrate the impact of semantic component grouping on improving code structure and quality, we conducted a user experiment inspired by the approach taken by Chen et al. \cite{chen2023egfe}. Specifically, we engaged two engineers with more than three years of experience in developing frontend code using JavaScript frameworks. Their assignment involved refining auto-generated code for 10 randomly selected pages from apps in shopping, music, and travel scenarios, adjusting the code to meet business needs. Using the Imgcook for code generation from design prototypes, we generated code for each UI page twice: once as original output and once after adjustments using semantic component grouping. The code for all pages was output in the React framework. We utilized code availability as a metric to indicate the enhancement of UI code structure and quality through semantic component grouping. It is defined as 
\begin{equation}
\begin{aligned}
   code\ availability = 1-\frac{lines\ of\ code\ changes}{total\ lines\ of\ code}.
\end{aligned}
\end{equation} 
Following the approach of Chen et al. \cite{chen2023egfe}, for the original scores ranging from 0 to 1, we applied a mapping based on the intervals [0.80, 0.85, 0.90, 0.95], converting them into a scale from 1 to 5. For the results, the code availability score for the original auto-generated code was 3.14. After optimization using semantic component grouping, the score for the generated code improved to 4.21, with a statistical significance of $p = 0.024^{*}$. This implies that semantic component grouping provides effective structural information for the code auto-generation process, allowing the generated code to be deployed with fewer adjustments required for practical use.

\subsection{Generate Accessibility Metadata for Screen Reader}
\label{Screen Reader}
Screen readers, as an assisted tool for visually impaired people to access UI without barriers, require the availability of accessibility data to support their services. However, it is still an important but often overlooked topic in UI design and implementation. Most of the apps in both the apple store and the android platform still do not universally supply accessibility information \cite{Screen_Recognition,49,50}, in which case the use of screen readers is challenged. For a specific UI element, the basic accessibility metadata includes features such as position, size, and semantic description. For example, for the ``watch'' icon in Fig. \ref{fig1}(a), the position is recorded as a quadruple form $[x_1,y_1,x_2,y_2]$, and the size as $[w,h]$. Semantic description denotes the icon's meaning or function, i.e., a watch. 

\begin{figure}[tbh]
    \centering
    \includegraphics[width=0.7\linewidth]{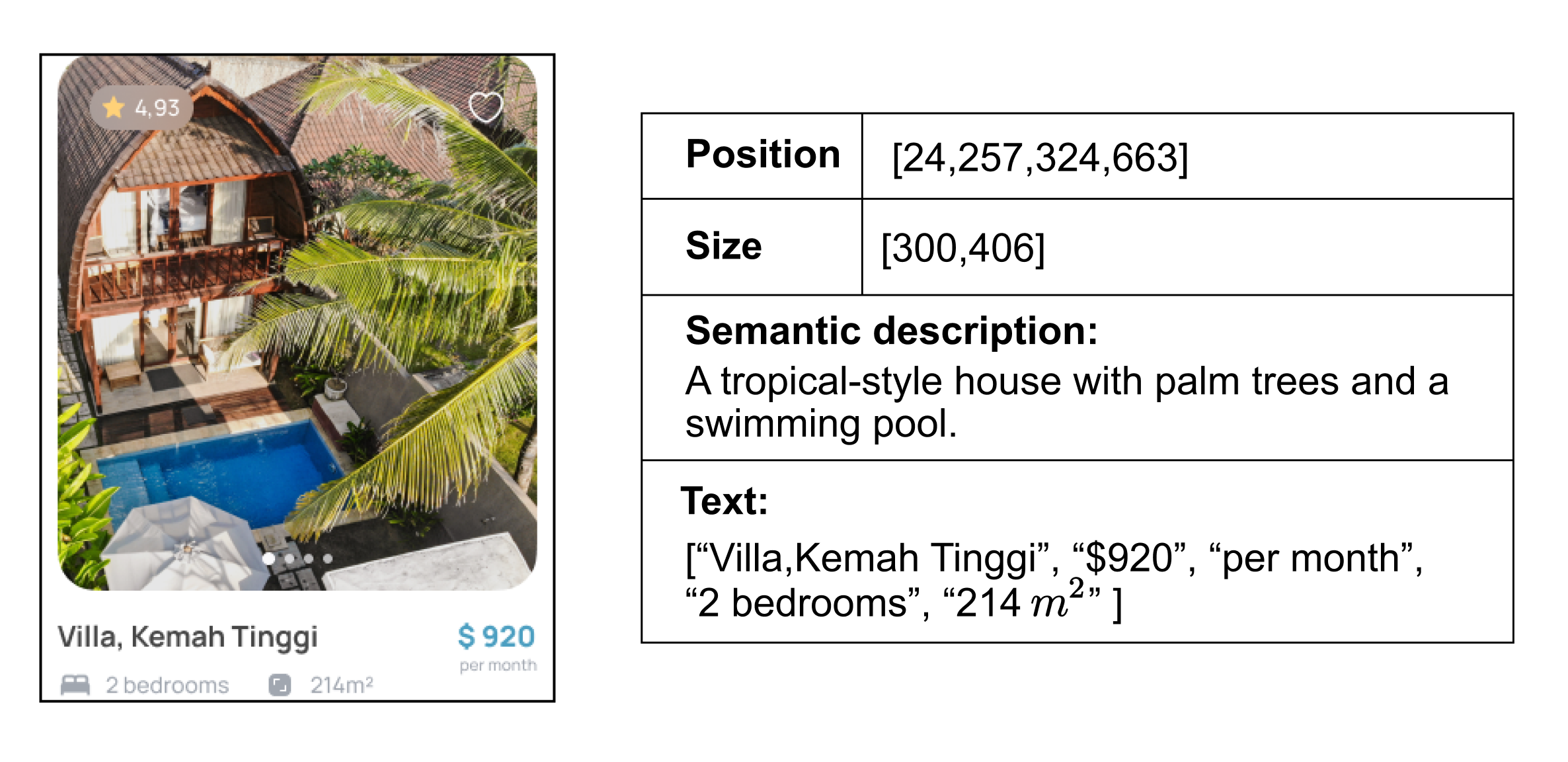}
    \caption{Example of accessibility data generated for our semantic component group. }
    \label{fig9}
    \vspace{-0.1in}
\end{figure}

To generate necessary accessibility metadata for screen readers, screen recognition proposed an on-device method mainly based on a UI widget recognition model. However, this approach only supports Apple apps and utilizes some built-in iOS features to help with generating the attributes we described above. In this section, we shortly present the idea of generating accessibility metadata for cross-platform apps by our semantic component groups. Given a UI screenshot, we perform accessibility data for each semantic component group, as shown in Fig. 9. The position and size directly come from the bounding boxes predicted by our UISCGD. For semantic description, we utilize the image captioning model \cite{52,53} to generate the image's content in words. And for icon groups, the icon recognition model \cite{icon_annotation} can be utilized, and we use the icon type for semantic description. Moreover, we record all text inside each semantic component group by utilizing open-source OCR tools \cite{13}.

\section{Discussion}
In the previous section, we first presented the detection performance of semantic component groups. Following that, we detailed how these groups can be applied to the automatic inference of section-level perceptual groups, their role in optimizing the structure and quality of frontend code through the automatic generation from UI design prototypes to code, and their utility in generating accessibility data required by screen readers. In this section, we summarize our findings and discuss potential limitations.

In the detection of semantic component groups, our method achieved an F1 score of 77.5\%, surpassing the second-best baseline method by 5.1\%. Compared to other real-world detection tasks, the detection of UI semantic component groups faces additional challenges involving targets of smaller size and those that are either narrow or elongated. These conditions pose challenges for deep visual algorithms in allocating attention effectively, as it becomes more difficult for sampling points to fall within the target range. To address this challenge, we introduced two strategies: colormap and prior group distribution. The former integrates the positional information of individual UI elements into the overall UI feature map, increasing the weight on potential grouping positions to force more sampling points to concentrate in these areas. The latter models features such as the position and aspect ratio of UI elements, implementing different sampling strategies for groups located in various positions on the UI interface. As a result, we observed that they respectively contributed to performance improvements of 3.3\% and 3.4\%. These two strategies also come with certain limitations. For the colormap strategy, its efficacy is impacted by the accuracy of the underlying single element detection algorithm, which in our case is the UIED method \cite{uied}. Particularly, false negative results from UIED may lead to some groups being overlooked. In UI interfaces, several factors such as the size of elements, image resolution, and the color contrast between elements and their background can lead to errors in UIED's detection performance. The prior group distribution strategy is modeled on the impact of a group's vertical positioning on its shape distribution within the UI page. This means the current scroll position of a screenshot can influence the effectiveness of this strategy. If the page is scrolled, changing the elements' positions relative to the viewport, it might alter the expected distribution of shapes, potentially affecting the accuracy of grouping based on this strategy. In this case, this strategy is particularly effective in identifying groups at the top (such as status bars) and bottom (like toolbars) of screenshots, as these groups remain in fixed positions even when the screen is scrolled.

In the inference of section-level perceptual groups, we compared our results with the method employed by Xie et al. \cite{perceptual_group}, which is based on Gestalt principles, and achieved an 8.3\% improvement in F1 score. In their method, the effectiveness of grouping is largely influenced by the accuracy of single element detection. Misidentification of text and images, as shown in Fig. \ref{fig6} is a primary issue that leads to the failure of subsequent processes based on similarity and proximity. In our approach, the holistic detection of semantic component groups, which include both image and text elements, eliminates this issue. However, our method is also influenced by the effectiveness of the semantic component group detection, which can be divided into two main aspects: In considerations based on the principle of similarity, since our algorithm identifies groups with similar shapes as a single perceptual group, fluctuations in the bounding boxes of predicted semantic component groups can lead to false negative issues. Regarding proximity, if not all semantic component groups that form a perceptual group are detected, this might result in incomplete recognition of the perceptual group or its segmentation into separate parts.

Regarding the optimization of automatically generated code through semantic component groups, we demonstrated visual improvements in the DOM tree by incorporating grouping constraints to mitigate the impact of incorrect view hierarchy arrangements in design prototypes on the code structure. Overall, the modified code structure aligns more closely with the visual structure and becomes more modular. Our user study also indicates that the modified code requires less manual intervention to meet business requirements. However, it's important to note that our approach primarily facilitates optimization at the component-level of code structure. It does not address issues related to finer-grained, fragmented layers or larger granularity grouping challenges. Additionally, our method is principally implemented based on the auto-generation logic of Imgcook. For other frontend code generation platforms, such as CodeFun \cite{codefun}, additional adjustments may be necessary.

Finally, we briefly presented how the results of semantic component grouping can aid in generating accessibility data. Since we did not actually integrate our grouping model with real-world accessibility tools, our description of how to implement this process was merely conceptual. Consequently, we did not conduct user validation similar to what was done with Screen Recognition \cite{Screen_Recognition}. In contrast to Screen Recognition, which necessitates initial detection of individual elements followed by rule-based grouping for accessibility data generation, our approach leverages the inherent semantic consistency or complementarity of elements within each semantic component group. This allows for the direct application of text and image understanding technologies to generate accessibility data, simplifying the process and potentially improving the efficiency and accuracy of accessibility feature development.

\section{Conclusion}
\label{sec:conclusion}


In this article, we propose a grouping method based on the semantic relevance of UI image and text elements, termed as semantic component groups. To infer these groups, we collected 1988 real-world mobile GUIs and constructed the UI semantic component group dataset through manual annotation. Our semantic component group detector, UISCGD, is built upon deformable-DETR and incorporates two strategies, colormap and prior group distribution, outperforming other SOTA object detectors by 6.1\% and achieving an F1 score of 77.5\%. Unlike other UI-related engineering tasks that rely on individual element detection and task-specific grouping rules, our approach captures groups directly, enabling application across multiple tasks. In this paper, we discuss the application of semantic component groups in three tasks: UI perceptual group partitioning, code structure improvement, and accessibility metadata generation. For UI perceptual group partitioning, our method achieves an F1 score 8.3\% higher than the psychologically-inspired approach, allowing for a more accurate understanding of section-level UI structure. For automatic code generation, we visualize the structure loss and achieve intuitive structure improvement. Our user study indicates that the modified code can meet business requirements with fewer modifications. For accessibility metadata, we demonstrate a simple procedure to generate necessary features for inaccessible apps. To further improve our work, structural information from GUI design prototypes can be utilized for better semantic component group detection performance in the future.

\bibliographystyle{ACM-Reference-Format}
\bibliography{bibliography}
\citestyle{acmauthoryear}
\end{document}